\documentclass[prd,twocolumn,showpacs,preprintnumbers,amsmath,amssymb,superscriptaddress,floatfix,nofootinbib]{revtex4-2}

\usepackage{graphicx}
\usepackage{bm}
\usepackage{amsmath}
\usepackage{amsfonts}
\usepackage{amssymb}
\usepackage{color}
\usepackage{multirow}
\usepackage{subfigure}
\usepackage{txfonts}
\usepackage[colorlinks, citecolor=blue,anchorcolor=red,menucolor=red, linkcolor=red,filecolor=red,runcolor=red,urlcolor=blue,frenchlinks=red]{hyperref}

\begin{document}
	\title{Explore the low-lying light baryons in  $\chi_{c0}\to pK^-\bar{\Lambda}\eta$ reaction}
		
	\author{Yun-He Lyu}
	\affiliation{School of Physics, Zhengzhou University, Zhengzhou 450001, China}
	\vspace{0.5cm}
		
    \author{Wen-Tao Lyu}\email{lvwentao9712@163.com}
    \affiliation{School of Physics, Zhengzhou University, Zhengzhou 450001, China}
    \affiliation{Departamento de Física Teórica and IFIC, Centro Mixto Universidad de Valencia-CSIC Institutos de Investigación de Paterna, 46071 Valencia, Spain}
    \vspace{0.5cm}
    
    \author{De-Min Li}\email{lidm@zzu.edu.cn}
    \affiliation{School of Physics, Zhengzhou University, Zhengzhou 450001, China}\vspace{0.5cm}
    
    \author{En Wang}\email{wangen@zzu.edu.cn}
    \affiliation{School of Physics, Zhengzhou University, Zhengzhou 450001, China}\vspace{0.5cm}

\begin{abstract}
Motivated by the recent BESIII measurements, we investigate the $\chi_{c0}\to pK^-\bar{\Lambda}\eta$ decay within the chiral unitary approach, incorporating $S$-wave meson-baryon final-state interactions. The coupled-channel amplitudes dynamically generate the $\Lambda(1405)$, $\Lambda(1670)$, and $N(1535)$ resonances, and an additional contribution from the intermediate $\Lambda(1520)$ state is included explicitly. A fit with six free parameters provides a reasonable overall description of all six measured two-body invariant mass distributions. Within this framework, the $\Lambda(1520)$ accounts for the prominent peak in the $pK^-$ distribution, while the final-state interactions produce resonance-related enhancements and threshold effects in the other subsystems. In particular, the $\bar{\Lambda}(1670)$ gives rise to a pronounced near-threshold enhancement in the $\eta\bar{\Lambda}$ channel. We estimate the theoretical uncertainties of the invariant mass distributions using a parametric bootstrap~(resampling) procedure. Higher-precision measurements of this decay would help constrain the final-state interactions and clarify the properties of the low-lying baryon resonances.

\end{abstract}
	
	\date{\today}
	
	\maketitle
	
\section{Introduction}\label{sec1}

The observation of the $X(3872)$ by the Belle Collaboration in 2003 stimulated extensive interest in exotic hadrons and the internal structure of hadronic states~\cite{Brambilla:2019esw,Chen:2022asf,Meng:2022ozq,Guo:2023xyf,Liu:2024uxn,Bai:2026atm,Lu:2024dtb}. In the light-baryon sector, the spectrum of the low-lying excited states also presents several longstanding puzzles. One example is the mass ordering of the $N(1535)$ with $J^P=1/2^-$ and the Roper resonance $N(1440)$ with $J^P=1/2^+$. In simple quark models, the lowest orbital excitation is expected to lie below the first radial excitation, whereas the observed ordering is reversed~\cite{ParticleDataGroup:2026mpi}. Such features have motivated continued investigations of the structure of the low-lying baryon resonances~\cite{Crede:2013kia}, which also contribute to our understanding of the nonperturbative dynamics of quantum chromodynamics (QCD) at low energies~\cite{Wang:2024jyk}.

Besides this mass-ordering puzzle, the $N(1535)$ exhibits strong couplings to the $\eta N$ and $K\Lambda$ channels, providing further motivation to investigate its internal structure~\cite{Liu:2005pm,Xie:2017erh,Pavao:2018wdf}. Various interpretations have been proposed, including a three-quark excitation with hidden-strangeness five-quark admixtures~\cite{Helminen:2000jb,Hannelius:2000gu,Zhang:2004xt,Zou:2007mk}, a resonance dynamically generated from meson--baryon interactions~\cite{Kaiser:1995eg,Kaiser:1996js,Inoue:2001ip,Nieves:2011gb}, and a three-quark core dressed by meson--baryon contributions~\cite{Liu:2015ktc,Guo:2022hud,Abell:2023nex}. Its properties have also been investigated in numerous other theoretical studies~\cite{Lyu:2023aqn,Li:2025gvo,Li:2026lbo,Li:2026umb,Li:2026wbl}. Despite these efforts, the relative importance of the different components remains an open question, and further studies are needed to clarify its nature.

Another resonance of interest is the $\Lambda(1670)$, which also has the spin-parity assignment $J^P=1/2^-$. It manifests as an enhancement in the distributions measured in the $K^-p\to\eta\Lambda$~~\cite{CrystalBall:2001uhc} and $\Lambda_c^+\to\eta\Lambda\pi^+$ processes~\cite{BESIII:2018qyg,Belle:2020xku,BESIII:2024mbf}, while a dip near the $\eta\Lambda$ threshold is observed in $K^-p\to\bar K^0n$~\cite{Gopal:1976gs}. These different line shapes can be influenced by threshold effects and interference between resonant and other contributions~\cite{Guo:2019twa,Wang:2022nac,Lyu:2024qgc,Duan:2024czu,Lyu:2026ack}. On the theoretical side, the $\Lambda(1670)$ has been studied both as a three-quark excitation~\cite{Zhong:2008km} and as a dynamically generated meson--baryon resonance~\cite{Oset:2001cn,Garcia-Recio:2002yxy,Oller:2006jw}. A substantial body of theoretical work has also explored its properties in different reactions and frameworks~\cite{Lyu:2024qgc,Duan:2024czu,Zhang:2024jby,Dai:2026zqn,Lyu:2026ack,Zhang:2026oml}. Its proximity to the $\eta\Lambda$ threshold and its reaction-dependent manifestations make it an interesting subject for further investigation.

Further insight into these resonances can be gained from their production in charmonium decays, where final-state interactions provide access to the dynamics of the relevant hadron pairs. Such decays have motivated numerous theoretical investigations of exotic hadrons and their production mechanisms~\cite{Wang:2020wap,Ding:2023eps,Ding:2024lqk,Lyu:2026ums,Dai:2026zqn,He:2026mkf}. In addition to three-body decay modes, a variety of four-body final states have been measured in particle decays~\cite{BESIII:2017jyh,Belle:2022ywa,BESIII:2026qzk,BESIII:2026gro}. The additional kinematic degrees of freedom, together with the possibility of interactions in several two-body subsystems, complicate the construction of a theoretical description. Theoretical analyses of these processes remain relatively limited. Studies of four-body decays have addressed $D_s^+\to\pi^+\pi^+\pi^-\eta$~\cite{Song:2022kac} and $D_s^+\to\pi^+\pi^0\pi^0\eta$~\cite{He:2026iiy}, as well as $\Lambda_c^+\to\Lambda\pi^+\pi^+\pi^-$~\cite{He:2025vij,Li:2025yad}, $D^+\to\bar{K}^0\pi^+\pi^0\pi^0$~\cite{Yang:2026bhm}, and $\psi(3686)\to p K^-\bar{K}^0\bar{\Xi}^0$~\cite{CesarSobrinho:2026vhe}.

 Using a sample of $(2712.4\pm14.3)\times10^6$ $\psi(3686)$ events, the BESIII Collaboration recently reported the first observation of $\chi_{cJ}\to pK^-\bar{\Lambda}\eta+\mathrm{c.c.}$ ($J=0,1,2$) and measured the corresponding branching fractions~\cite{BESIII:2026gro}. This four-body final state contains six two-body subsystems and thus offers more opportunities to investigate final-state interactions than a three-body final state. The corresponding invariant mass distributions provide complementary information on potential resonance contributions and the dynamics of the interacting hadron pairs. In particular, the measured invariant mass distributions exhibit a pronounced peak near 1520 MeV in the $pK^-$ mass distribution, an enhancement around 1535 MeV in the $p\eta$ distribution, and a near-threshold enhancement in the $\eta\bar{\Lambda}$ distribution. The first two features occur in the mass regions of the $\Lambda(1520)$ and $N(1535)$ resonances, respectively, while the near-threshold enhancement may indicate significant final-state interactions in the $\eta\bar{\Lambda}$ system. These structures motivate a theoretical investigation of the roles of intermediate resonances and final-state interactions in this decay. Although the multidimensional phase space and the possible interplay among different final-state interactions complicate its theoretical description, this process provides a promising setting for studying baryon resonances.

In the present work, we study the decay $\chi_{c0}\to pK^-\bar{\Lambda}\eta$, treating the final-state interactions~(FSIs) of the $K^-p$, $\eta\bar{\Lambda}$, and $\eta p$ pairs within the chiral unitary approach. In this framework, the coupled-channel meson--baryon amplitudes dynamically generate the two poles associated with the $\Lambda(1405)$, together with the $\Lambda(1670)$ resonance in the isospin-zero sector and the $N(1535)$ resonance in the isospin-half sector. We analyze how these resonances shape the two-body invariant mass spectra and account for an additional $\Lambda(1520)$ contribution in the $K^-p$ channel. By fitting the available experimental data, we determine the model parameters and obtain the invariant mass distributions for all six two-body subsystems, allowing us to assess the agreement with the measured spectra and investigate the resonance contributions.

The remainder of this paper is organized as follows. In Sec.~\ref{sec2}, we present the theoretical formalism for the $\chi_{c0}\to pK^-\bar{\Lambda}\eta$ decay. The numerical results and their discussion are presented in Sec.~\ref{sec3}. Finally, our conclusions are given in Sec.~\ref{sec4}.

\section{Formalism}\label{sec2}

\subsection{Final-state interaction contributions}

\begin{figure}[htbp]
	\centering
	
	\includegraphics[scale=0.45]{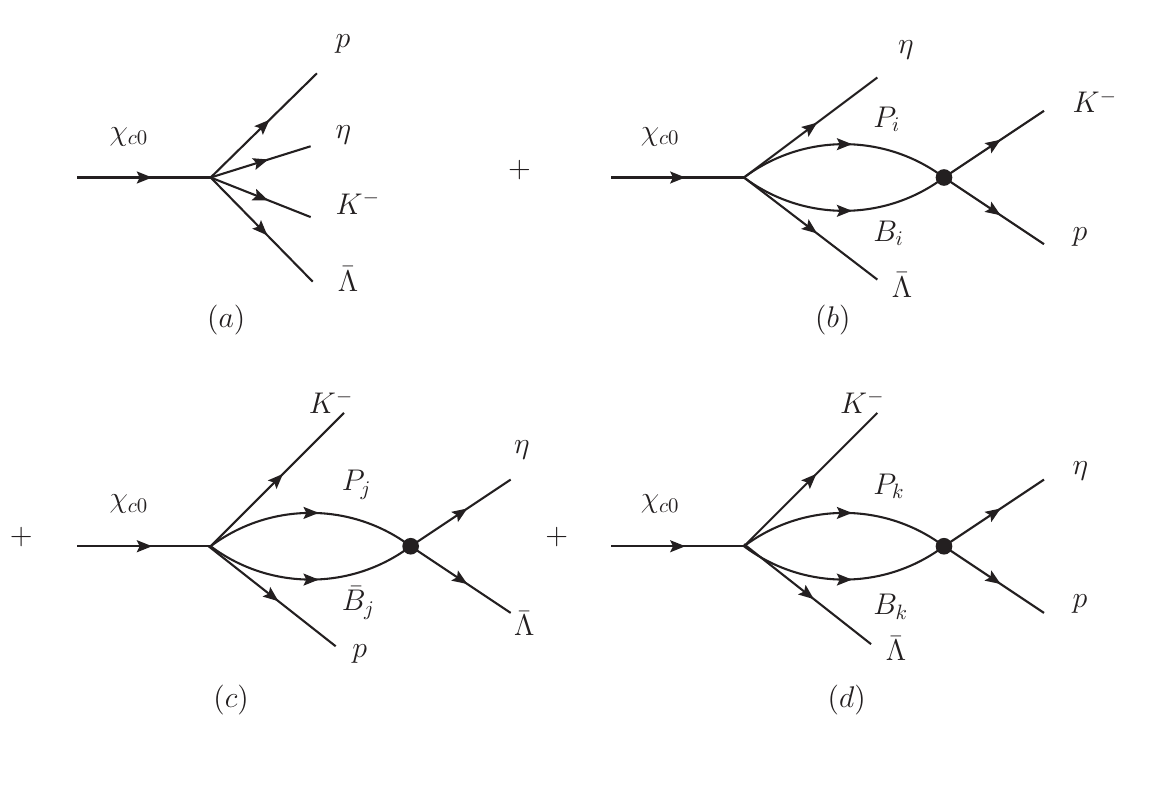}
	
	\caption{Mechanisms for the tree-level $\chi_{c0}\to pK^-\bar{\Lambda}\eta$ decay and subsequent rescattering of intermediate hadron pairs.}\label{fig:chic0_decay}
\end{figure}

We begin by considering the strong decay $\chi_{c0}\to pK^-\bar{\Lambda}\eta$. In this work, we account for FSIs in the $K^-p$ and $\eta p$ meson--baryon subsystems, and the $\eta\bar{\Lambda}$ meson--antibaryon subsystems. The relevant hadronic channels are constructed using the pseudoscalar meson, baryon, and antibaryon octet matrices, denoted by $P$, $B$, and $\bar{B}$, respectively.

To identify the meson--baryon and meson--antibaryon channels produced in this decay, we treat $\chi_{c0}$ as an SU(3) singlet in the $u$, $d$, $s$ sector. We express the $q\bar{q}$, $qqq$, and $\bar{q}\bar{q}\bar{q}$ states in terms of the meson, baryon, and antibaryon octet matrices, respectively. Using the $\eta$--$\eta^\prime$ mixing scheme of Ref.~\cite{Bramon:1992kr}, we obtain

\begin{eqnarray}\label{Eq:P_matrix}
	P=
	\left(
	\begin{array}{ccc}
		\frac{1}{\sqrt{2}}\pi^0 + \frac{1}{\sqrt{3}} \eta & \pi^+ & K^+ \\[2mm]
		\pi^- & -\frac{1}{\sqrt{2}} \pi^0 + \frac{1}{\sqrt{3}} \eta & K^0 \\[2mm]
		K^- & \bar{K}^0 & ~-\frac{1}{\sqrt{3}} \eta\\
	\end{array}
	\right), 
\end{eqnarray}
\begin{eqnarray}
	\label{Eq:B_matrix}
	B=
	\left(
	\begin{array}{ccc}
		\frac{1}{\sqrt{2}}\Sigma^0 + \frac{1}{\sqrt{6}} \Lambda & \Sigma^+ & p\\[2mm]
		\Sigma^- & -\frac{1}{\sqrt{2}} \Sigma^0 + \frac{1}{\sqrt{6}} \Lambda & n \\[2mm]
		\Xi^- & \Xi^0 & ~-\frac{2}{\sqrt{6}} \Lambda\\
	\end{array}
	\right) ,
\end{eqnarray}
\begin{eqnarray}
	\label{Eq:Bbar_matrix}
	\bar{B}=
	\left(
	\begin{array}{ccc}
		\frac{1}{\sqrt{2}}\bar{\Sigma}^0 + \frac{1}{\sqrt{6}} \bar{\Lambda} & \bar{\Sigma}^+ & \bar{\Xi}^+ \\[2mm]
		\bar{\Sigma}^- & -\frac{1}{\sqrt{2}} \bar{\Sigma}^0 + \frac{1}{\sqrt{6}} \bar{\Lambda }&  \bar{\Xi}^0\\[2mm]
		\bar{p} & \bar{n} & ~-\frac{2}{\sqrt{6}} \bar{\Lambda}\\
	\end{array}
	\right).
\end{eqnarray}

We consider the SU(3)-singlet structures formed by taking traces of products of these matrices, specifically $\langle\bar{B}PPB\rangle$, $\langle\bar{B}PBP\rangle$, and $\langle\bar{B}BPP\rangle$. Arguments based on heavy-quark symmetry indicate that structures with fewer traces provide the dominant contributions~\cite{Manohar:1998xv,Abreu:2023yvf}. We therefore restrict our analysis to these three single-trace structures.

Accordingly, we parameterize the SU(3)-invariant amplitude as
\begin{equation}\label{Eq:Trace}
	\mathcal{M}= A\,\mathrm{Tr}(\bar{B}PPB) + B\,\mathrm{Tr}(\bar{B}PBP) + C\,\mathrm{Tr}(\bar{B}BPP),
\end{equation}
where $A$, $B$, and $C$ are the coefficients of the three independent trace structures.

Expanding the traces in terms of physical meson, baryon, and antibaryon fields, we group the resulting contributions according to the selected two-body subsystem. For the $\eta\bar{\Lambda}$ pair, the contribution involving the accompanying meson--baryon channels reads
\begin{equation}\label{Eq:H_etaLambda}
\begin{aligned}
	\mathcal{M}_{\eta\bar{\Lambda}}
	= \eta\bar{\Lambda}\Bigg\{&
	\frac{B}{\sqrt{2}}
	\left(K^-p+\bar{K}^{0}n+K^+\Xi^-+K^{0}\Xi^{0}\right)
	\\
	&+\frac{\sqrt{2}}{3}(A+B+C)
	\left(\pi^{0}\Sigma^{0}+\pi^{+}\Sigma^{-}+\pi^{-}\Sigma^{+}\right)
	\\
	&+\frac{1}{3}(A+B+C)\eta\Lambda
	\Bigg\}.
\end{aligned}
\end{equation}

For the $K^-p$ pair, the corresponding meson--antibaryon contributions are
\begin{equation}\label{Eq:H_Kp}
\begin{aligned}
	\mathcal{M}_{K^-p}
	=K^-p\Bigg\{&
	(A+C)K^+\bar{p}+A K^{0}\bar{n}
	-\frac{2A-B-C}{2\sqrt{3}}\pi^{0}\bar{\Lambda}
	\\
	&+\frac{B+C}{2}\pi^{0}\bar{\Sigma}^{0}
	+\frac{B}{\sqrt{2}}\eta\bar{\Lambda}
	+\frac{B}{\sqrt{6}}\eta\bar{\Sigma}^{0}
	\\
	&+B\pi^-\bar{\Sigma}^{+}
	+C\pi^+\bar{\Sigma}^{-}
	+B K^-\bar{\Xi}^{+}
	+B\bar{K}^{0}\bar{\Xi}^{0}
	\Bigg\}.
\end{aligned}
\end{equation}
Finally, selecting the $K^-\bar{\Lambda}$ pair instead gives the following contributions from the accompanying meson--baryon channels:
\begin{equation}\label{Eq:H_KLambda}
	\begin{aligned}
		\mathcal{M}_{K^-\bar{\Lambda}}
		=K^-\bar{\Lambda}\Bigg\{&
		\frac{A-2B+C}{\sqrt{6}}K^{0}\Sigma^{+}
		\\
		&+\frac{A-2B+C}{2\sqrt{3}}K^{+}\Sigma^{0}
		\\
		&+\frac{5A-4B+5C}{6}K^{+}\Lambda
		\\
		&-\frac{2A-B-C}{\sqrt{6}}\pi^{+}n
		\\
		&-\frac{2A-B-C}{2\sqrt{3}}\pi^{0}p
		+\frac{B}{\sqrt{2}}\eta p
		\Bigg\}.
	\end{aligned}
\end{equation}

The hadronization expressions above determine the flavor coefficients of the relevant channels containing two mesons, a baryon, and an antibaryon. We now construct the primary production vertex and incorporate the subsequent final-state interactions. We take the spinor structure of the primary vertex to be
\begin{equation}
	t=\bar{u}(p)v(p'),
\end{equation}
where $p$ and $p'$ are the four‑momenta of the baryon and antibaryon. Retaining terms up to first order in the momenta, we use the spinors
\begin{equation}
	u_B(p) = \begin{pmatrix}
		\chi_r  \\[8pt]
		\frac{\vec{\sigma}\cdot\vec{p}}{2m}\chi_r  
	\end{pmatrix},~~~~~~v_{B^\prime}(p^\prime) = \begin{pmatrix}
		\frac{\vec{\sigma}\cdot\vec{p^\prime}}{2m^\prime}\chi_r^\prime  \\[8pt]
		\chi_r^\prime  
	\end{pmatrix}.
\end{equation}
Using this vertex, we construct the transition amplitude for the $\chi_{c0}\to pK^-\bar{\Lambda}\eta$ decay from the mechanisms shown in Fig.~\ref{fig:chic0_decay}. In the following expressions, all spatial components are treated as contravariant, and repeated spatial indices are summed over. The spinor structure of the tree-level vertex, shown in the first diagram of Fig.~\ref{fig:chic0_decay}, reduces to
\begin{align}
	\bar{u}(p)v(p') \to 
	\frac{\vec{\sigma}\cdot\vec{p^\prime}}{2m^\prime} - \frac{\vec{\sigma}\cdot\vec{p}}{2m}.
\end{align}
Both terms contribute to the tree-level amplitude. In the rescattering diagrams, however, terms proportional to the momentum of the intermediate baryon in the loop vanish because the scattering amplitudes are evaluated in the $S$-wave~\cite{Ikeno:2026zpv,Lyu:2026rsm,Ren:2026ila,Song:2026gnn}. The total transition amplitude for the mechanisms shown in Fig.~\ref{fig:chic0_decay} can therefore be written as
\begin{equation}
	\begin{aligned}
	\mathcal{M}_{FSI}=&\frac{\vec{\sigma}\cdot\vec{p^\prime}}{2m^\prime}\times\left(t_{\text{tree}}+t_{k^-p}+t_{\eta p}\right) \\
	&- \frac{\vec{\sigma}\cdot\vec{p }}{2m}\times\left(t_{\text{tree}}+t_{\eta\bar{\Lambda}}\right),
	\end{aligned}
\end{equation}
where 
\begin{equation}\label{Eq:t_tree}
	t_{\text{tree}}=\frac{B}{\sqrt{2}},
\end{equation}
\begin{equation}\label{Eq:t_k-p}
	\begin{aligned}
    t_{k^-p}=&\sum_ih_iG_{P_iB_i}t_{P_iB_i,K^-p}\left(M_{K^-p}\right)\\
    =&\frac{B}{\sqrt{2}}G_{K^-p}
    t_{K^-p,K^-p}+\frac{B}{\sqrt{2}}G_{\bar{K}^{0}n}
    t_{\bar{K}^{0}n,K^-p}
    \\
    &+\frac{\sqrt{2}}{3}(A+B+C)G_{\pi^{0}\Sigma^{0}}
    t_{\pi^{0}\Sigma^{0},K^-p}
    \\
    &+\frac{2}{3}(A+B+C)G_{\eta\Lambda}
    t_{\eta\Lambda,K^-p}
    \\
    &+\frac{\sqrt{2}}{3}(A+B+C)G_{\pi^{+}\Sigma^{-}}
    t_{\pi^{+}\Sigma^{-},K^-p}
    \\
    &+\frac{\sqrt{2}}{3}(A+B+C)G_{\pi^{-}\Sigma^{+}}
    t_{\pi^{-}\Sigma^{+},K^-p}
    \\
    &+\frac{B}{\sqrt{2}}G_{K^{+}\Xi^{-}}\,
    t_{K^{+}\Xi^{-},K^-p}+\frac{B}{\sqrt{2}}G_{K^{0}\Xi^{0}}\,
    t_{K^{0}\Xi^{0},K^-p},
    \end{aligned}
\end{equation}
\begin{equation}\label{Eq:t_etaLambdabar}
	\begin{aligned}
    t_{\eta\bar{\Lambda}}=&\sum_jh_jG_{P_j\bar{B}_j}t_{P_j\bar{B}_j,\eta\bar{\Lambda}}\left(M_{\eta \bar{\Lambda}}\right)\\
    =&(A+C)G_{K^{+}\bar{p}}\,
    t_{K^{+}\bar{p},\eta\bar{\Lambda}}
    +A G_{K^{0}\bar{n}}\,
    t_{K^{0}\bar{n},\eta\bar{\Lambda}}
    \\
    &+\frac{B+C-2A}{2\sqrt{3}}G_{\pi^{0}\bar{\Lambda}}\,
    t_{\pi^{0}\bar{\Lambda},\eta\bar{\Lambda}}
    \\
    &+\frac{B+C}{2}G_{\pi^{0}\bar{\Sigma}^{0}}\,
    t_{\pi^{0}\bar{\Sigma}^{0},\eta\bar{\Lambda}}
    \\
    &+\frac{B}{\sqrt{2}}G_{\eta\bar{\Lambda}}\,
    t_{\eta\bar{\Lambda},\eta\bar{\Lambda}}
    +\frac{B}{\sqrt{6}}G_{\eta\bar{\Sigma}^{0}}\,
    t_{\eta\bar{\Sigma}^{0},\eta\bar{\Lambda}}
    \\
    &+B G_{\pi^{-}\bar{\Sigma}^{+}}\,
    t_{\pi^{-}\bar{\Sigma}^{+},\eta\bar{\Lambda}}
    +C G_{\pi^{+}\bar{\Sigma}^{-}}\,
    t_{\pi^{+}\bar{\Sigma}^{-},\eta\bar{\Lambda}}
    \\
    &+2B G_{K^{-}\bar{\Xi}^{+}}\,
    t_{K^{-}\bar{\Xi}^{+},\eta\bar{\Lambda}}
    +B G_{\bar{K}^{0}\bar{\Xi}^{0}}\,
    t_{\bar{K}^{0}\bar{\Xi}^{0},\eta\bar{\Lambda}},
    \end{aligned}
\end{equation}
\begin{equation}\label{Eq:t_etap}
	\begin{aligned}
t_{\eta p}=&\sum_kh_k\tilde{G}_{P_kB_k}\tilde{t}_{P_kB_k,\eta p}\left(M_{\eta p}\right)
\\
=&\frac{B}{\sqrt{2}}\tilde{G}_{\eta p}\,
\tilde{t}_{\eta p,\eta p}
+\frac{A-2B+C}{\sqrt{6}}\tilde{G}_{K^{0}\Sigma^{+}}\,
\tilde{t}_{K^{0}\Sigma^{+},\eta p}
\\
&+\frac{A-2B+C}{2\sqrt{3}}\tilde{G}_{K^{+}\Sigma^{0}}\,
\tilde{t}_{K^{+}\Sigma^{0},\eta p}
\\
&+\frac{5A-4B+5C}{6}\tilde{G}_{K^{+}\Lambda}\,
\tilde{t}_{K^{+}\Lambda,\eta p}
\\
&-\frac{2A-B-C}{\sqrt{6}}\tilde{G}_{\pi^{+}n}\,
\tilde{t}_{\pi^{+}n,\eta p}
\\
&-\frac{2A-B-C}{2\sqrt{3}}\tilde{G}_{\pi^{0}p}\,
\tilde{t}_{\pi^{0}p,\eta p},
\end{aligned}
\end{equation}

The factors of 2 associated with identical particles have already been included in these expressions. We also use the relation $t_{M\bar{B},M\bar{B}}=t_{MB,MB}$ to express the meson--antibaryon amplitudes in terms of the available meson--baryon amplitudes.

For the scattering amplitudes $t_{i,K^-p}$ and $t_{j,\eta\bar{\Lambda}}$, we employ the chiral unitary approach described in Ref.~\cite{Oset:1997it}, with the following coupled channels: $K^-p$ (1), $\bar{K}^{0}n$ (2), $\pi^{0}\Lambda$ (3), $\pi^{0}\Sigma^{0}$ (4), $\eta\Lambda$ (5), $\eta\Sigma^{0}$ (6), $\pi^{+}\Sigma^{-}$ (7), $\pi^{-}\Sigma^{+}$ (8), $K^{+}\Xi^{-}$ (9), and $K^{0}\Xi^{0}$ (10). The scattering matrix is obtained from
\begin{equation}\label{Eq:BS}
	T = [1-VG]^{-1}V,
\end{equation}
where $G_i$ is the standard meson--baryon loop function. We evaluate the loop functions using a three--momentum cutoff method. The transition potential is given by
 \begin{align}\label{Eq:Vij}
 	V_{ij}=-\frac{C_{ij}}{4f_if_j}\left(k_i^0+k_j^0\right),
 \end{align}
with
\begin{equation}
	k_i^0(\sqrt{s})=\frac{s+m_i^2-M_i^2}{2\sqrt{s}}.
	\label{eq:meson-energy}
\end{equation}
Here, $m_i$ and $M_i$ denote the meson and baryon masses in channel $i$, respectively. The coefficients $C_{ij}$ are given in Ref.~\cite{Oset:1997it}, and the values of $q_{\text{max}}$ and $f_i$ are given in Ref.~\cite{Duan:2024okk}.

For the contribution $\tilde{G}_k\tilde{t}_{k,\eta p}$, we likewise employ the chiral unitary approach, with the coupled channels $K^{0}\Sigma^{+}$ (1), $K^{+}\Sigma^{0}$ (2), $K^{+}\Lambda$ (3), $\pi^{+}n$ (4), $\pi^{0}p$ (5), and $\eta p$ (6). The transition potential is given by
\begin{equation}
	V_{ij}
	=-\frac{\tilde{C}_{ij}}{4\tilde{f}_i\tilde{f}_j}\left(k_i^0+k_j^0\right),
	\label{eq:kernel}
\end{equation}
where
\begin{equation}
	\tilde{f}_\pi=93~\text{MeV},\qquad
	\tilde{f}_K=1.22f_\pi,\qquad
	\tilde{f}_\eta=1.30f_\pi.
	\label{eq:decay-constants}
\end{equation}
The coefficients $\tilde{C}_{ij}$ are listed in Table~\ref{Tab:Cij}. The loop functions $\tilde{G}$ are regularized using a three-momentum cutoff of $\tilde{q}_{\text{max}}=1000$~MeV~\cite{Song:2026gnn}.
\begin{table}
	\centering
	\caption{Coefficients $\tilde{C}_{ij}$ entering the interaction potential.}
	\label{Tab:Cij}
	$
	\begin{array}{c|cccccc}
		\hline\hline
		& K^0\Sigma^+ & K^+\Sigma^0 & K^+\Lambda
		& \pi^+n & \pi^0p & \eta p \\
		\hline
		K^0\Sigma^+ & 1 & \sqrt{2} & 0 & 0
		& 1/\sqrt{2} & -\sqrt{3/2} \\
		K^+\Sigma^0 & & 0 & 0 & 1/\sqrt{2}
		& -1/2 & -\sqrt{3}/2 \\
		K^+\Lambda & & & 0 & -\sqrt{3/2}
		& -\sqrt{3}/2 & -3/2 \\
		\pi^+n & & & & 1 & \sqrt{2} & 0 \\
		\pi^0p & & & & & 0 & 0 \\
		\eta p & & & & & & 0 \\
		\hline\hline
	\end{array}
	$
\end{table}

\subsection{Contribution of $\Lambda(1520)$ to $\chi_{c0}\to pK^-\bar{\Lambda}\eta$}

\begin{figure}[htbp]
	\centering
	\includegraphics[scale=0.45]{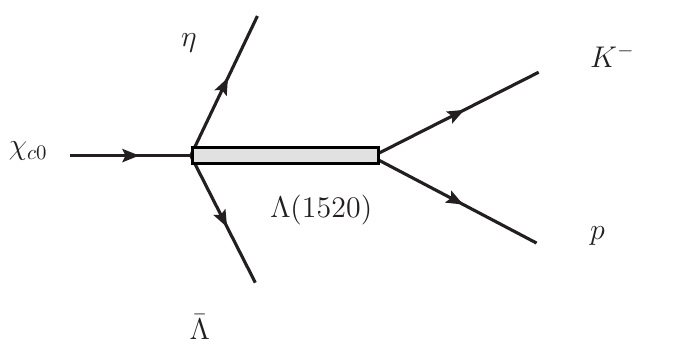}
	\caption{Mechanism for the decay $\chi_{c0}\to pK^-\bar{\Lambda}\eta$ via $\Lambda(1520)$ excitation.
	}\label{fig:Lambda1520}
\end{figure}

The mechanism involving the $\Lambda(1520)$ state with spin-parity $3/2^-$ is shown in Fig.~\ref{fig:Lambda1520}. Conservation of spin and parity requires $L=1$ for the coupling of $\chi_{c0}$ to $\eta\bar{\Lambda}\Lambda(1520)$ and $L=2$ for the coupling of $\Lambda(1520)$ to $K^-p$. We use an effective angle-averaged amplitude,
\begin{equation}\label{Eq:t1520}
	\mathcal{M}_{\Lambda(1520)}=\frac{D}{M_\Lambda^2}\dfrac{|\vec{P}_{\eta\bar{\Lambda}}||\vec{\tilde{P}}_{K^-}|^2}{M_{\text{inv}}(K^- p)-M_{\Lambda(1520)}+{i\Gamma_{\Lambda(1520)}}/{2}},  
\end{equation}
where $D$ is a free parameter. The factor $M_\Lambda^{-2}$ is included so that $D$ has the same dimensions as $A$, $B$, and $C$. In our calculation, we use $M_{\Lambda(1520)}=1519.42$~MeV and $\Gamma_{\Lambda(1520)}=15.80$~MeV. The three-momenta $\vec{P}_{\eta\bar{\Lambda}}$ and $\vec{\tilde{P}}_{K^-}$ refer to the $\eta\bar{\Lambda}$ system in the $\chi_{c0}$ rest frame and the $K^-$ meson in the $K^-p$ rest frame, respectively. Their magnitudes are given by
\begin{equation}
|\vec{P}_{\eta\bar{\Lambda}}|=\frac{\lambda^{1/2}\!\left(M_{\chi_{c0}}^2,M_{\eta\bar{\Lambda}}^2,M_{K^-p}^2\right)}{2M_{\chi_{c0}}},
\end{equation}
and
\begin{equation}
|\vec{\tilde{P}}_{K^-}|=\frac{\lambda^{1/2}\!\left(M_{K^-p}^2,M_K^2,M_p^2\right)}{2M_{K^-p}},
\end{equation}

\subsection{Full decay amplitude}

The full amplitude for the $\chi_{c0}\to pK^-\bar{\Lambda}\eta$ decay is then given by
\begin{equation}\label{Eq:ttotal}
	\mathcal{M}^{\text{total}}=\mathcal{M}_{FSI}+\mathcal{M}_{\Lambda(1520)}. 
\end{equation}
Neglecting interference between the $\Lambda(1520)$ contribution and the remaining terms, the squared amplitude, summed over final-state spins, takes the form
\begin{equation}
\begin{aligned}
	&\sum_{\mathrm{spins}}\left|\mathcal{M}^{\text{total}}\right|^2 \\
	={}&2\Bigg[
	\frac{|\vec{p}_{\bar{\Lambda}}|^{\,2}}{4M_{\bar{\Lambda}}^2}
	\Big(
	|t_{\text{tree}}|^2+|t_{k^-p}|^2+|t_{\eta p}|^2\\
	&+2\operatorname{Re}\!\left(t_{\text{tree}}^*t_{k^-p}\right)
	+2\operatorname{Re}\!\left(t_{\text{tree}}^*t_{\eta p}\right)
	+2\operatorname{Re}\!\left(t_{k^-p}^*t_{\eta p}\right)
	\Big)\\
	&+\frac{|\vec{p}_p|^{\,2}}{4M_p^2}
	\Big(
	|t_{\text{tree}}|^2+|t_{\eta\bar{\Lambda}}|^2
	+2\operatorname{Re}\!\left(t_{\text{tree}}^*t_{\eta\bar{\Lambda}}\right)
	\Big)\\
	&-\frac{\vec{p}_{\bar{\Lambda}}\cdot\vec{p}_p}
	{2M_{\bar{\Lambda}}M_p}
	\Big(
	|t_{\text{tree}}|^2
	+\operatorname{Re}\!\left(t_{\text{tree}}^*t_{\eta\bar{\Lambda}}\right)\\
	&+\operatorname{Re}\!\left(t_{k^-p}^*t_{\text{tree}}\right)
	+\operatorname{Re}\!\left(t_{k^-p}^*t_{\eta\bar{\Lambda}}\right)\\
	&+\operatorname{Re}\!\left(t_{\eta p}^*t_{\text{tree}}\right)
	+\operatorname{Re}\!\left(t_{\eta p}^*t_{\eta\bar{\Lambda}}\right)
	\Big)
	\Bigg]
	+|\mathcal{M}_{\Lambda(1520)}|^2.
\end{aligned}
\end{equation}
No average over initial-state spins is required because the parent meson has spin zero. The proton and antihyperon three-momenta are evaluated in the rest frame of the parent meson, with magnitudes
\begin{equation}
|\boldsymbol{p}_{\bar{\Lambda}}|=\frac{\lambda^{1/2}\!\left(M_{\chi_{c0}}^2,M_{\bar{\Lambda}}^2,M_{K^-\eta p}^2\right)}{2M_{\chi_{c0}}},
\end{equation}
and
\begin{equation}
|\vec{p}_p|=\frac{\lambda^{1/2}\!\left(M_{\chi_{c0}}^2,M_p^2,M_{K^-\eta\bar{\Lambda}}^2\right)}{2M_{\chi_{c0}}},
\end{equation}
where the Källén function is defined as
\begin{equation}
\lambda(x,y,z)=x^2+y^2+z^2-2xy-2xz-2yz.
\end{equation}
The scalar product of the two three-momenta can be expressed in terms of the proton--antihyperon invariant mass as
\begin{equation}
\vec{p}_p\cdot\vec{p}_{\bar{\Lambda}}=\frac{1}{2}\left(M_p^2+M_{\bar{\Lambda}}^2+2E_pE_{\bar{\Lambda}}-M_{\bar{\Lambda}p}^2\right),
\end{equation}
where the proton and antihyperon energies are given by
\begin{equation}
E_p=\frac{M_{\chi_{c0}}^2+M_p^2-M_{K^-\eta\bar{\Lambda}}^2}{2M_{\chi_{c0}}},
\end{equation}
and
\begin{equation}
E_{\bar{\Lambda}}=\frac{M_{\chi_{c0}}^2+M_{\bar{\Lambda}}^2-M_{K^-\eta p}^2}{2M_{\chi_{c0}}}.
\end{equation}
The invariant masses appearing in these expressions correspond to the final-state subsystems indicated by their subscripts.

Using the Mandl and Shaw normalization for the meson and baryon fields~\cite{Mandl:1985bg}, we obtain the differential decay width by combining the spin-summed squared amplitude with the Lorentz-invariant four-body phase-space measure:
\begin{equation}
d\Gamma=\frac{2M_{p}2M_{\bar{\Lambda}}}{2M_{\chi_{c0}}}\sum_{\mathrm{spins}}|\mathcal{M}^{\text{total}}|^2\,d\Phi_4,
\end{equation}
where the four-body phase-space measure is
\begin{equation}
d\Phi_4=(2\pi)^4\delta^{(4)}\left(P-\sum_{i=1}^{4}p_i\right)\prod_{i=1}^{4}\frac{d^3\boldsymbol{p}_i}{(2\pi)^3\,2E_i}.
\end{equation}
The symbol $P$ denotes the four-momentum of the parent meson, while $p_i$ and $E_i$ denote the four-momentum and energy of the final-state particle $i$, respectively.

The phase-space integration is performed numerically using the Julia package \texttt{DalitzPlot.jl}~\cite{DalitzPlot}. We generate $2\times10^8$ Monte Carlo events over the kinematically allowed four-body phase space, with all final-state particles on shell and total four-momentum conserved. For each event, the spin-summed squared amplitude is evaluated and multiplied by the corresponding Monte Carlo integration weight. The resulting weighted sample is used to calculate the invariant mass distributions of the selected final-state subsystems.


\section{Results and Discussions}\label{sec3}

\begin{figure*}[htbp]
	\centering
	
	\subfigure[]{
		\includegraphics[width=0.32\textwidth]{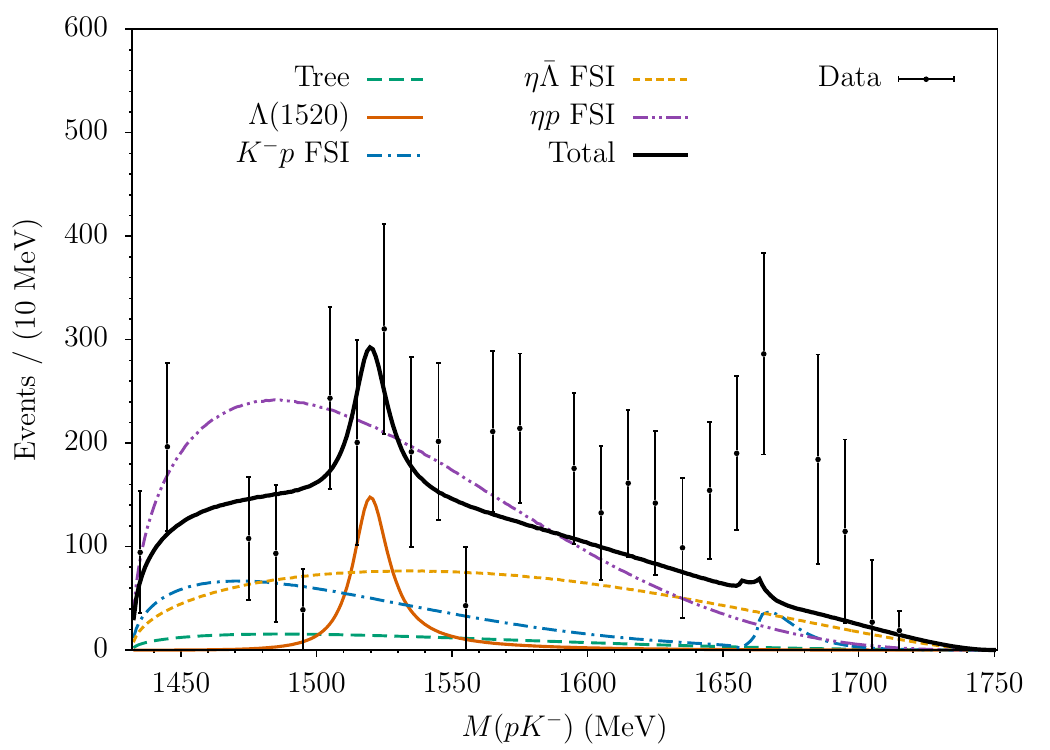}
		\label{fig:fit_pKm}
	}\hfill
	\subfigure[]{
		\includegraphics[width=0.32\textwidth]{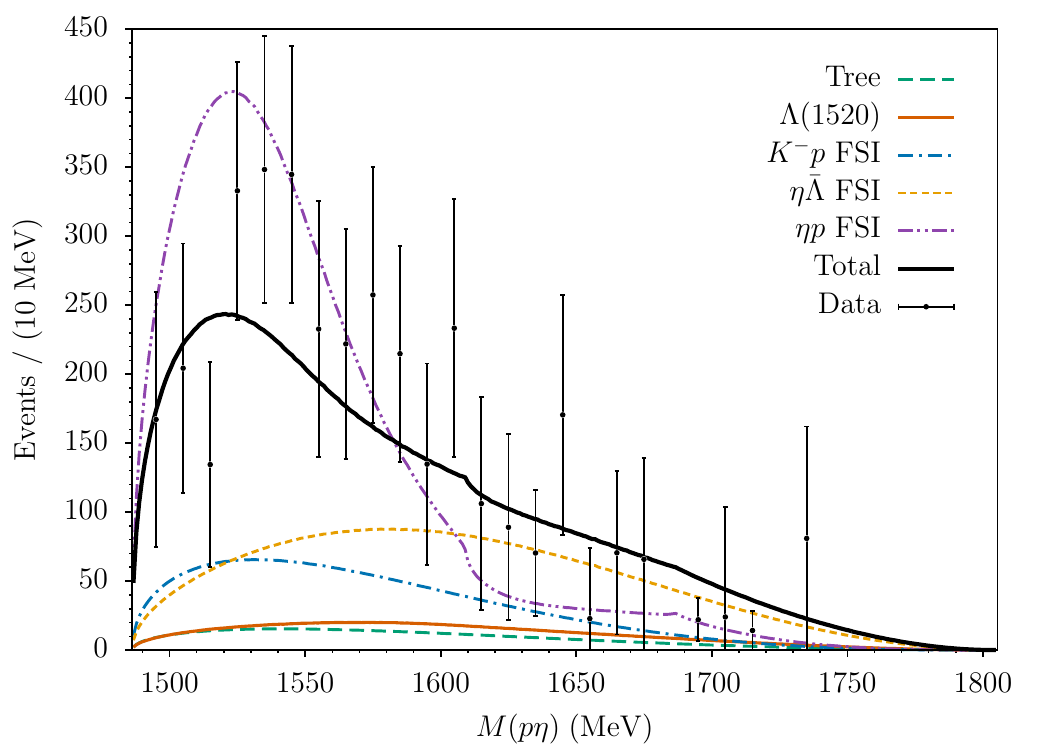}
		\label{fig:fit_peta}
	}\hfill
	\subfigure[]{
		\includegraphics[width=0.32\textwidth]{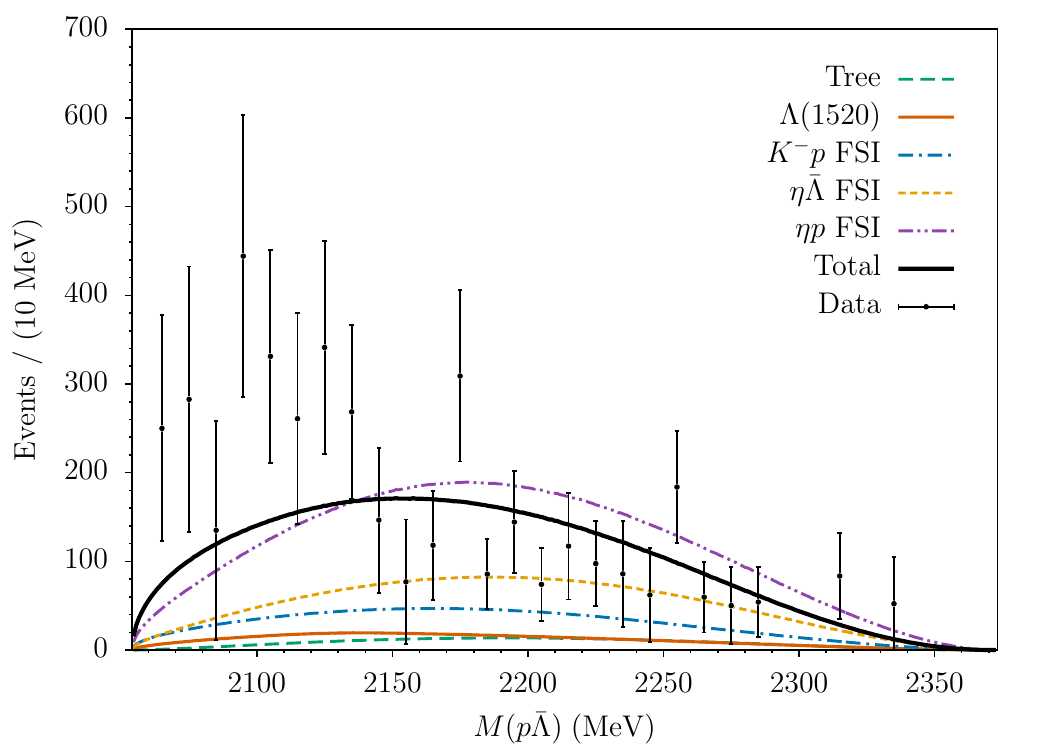}
		\label{fig:fit_pLambar}
	}
	
	\par\vspace{0.3cm}
	
	\subfigure[]{
		\includegraphics[width=0.32\textwidth]{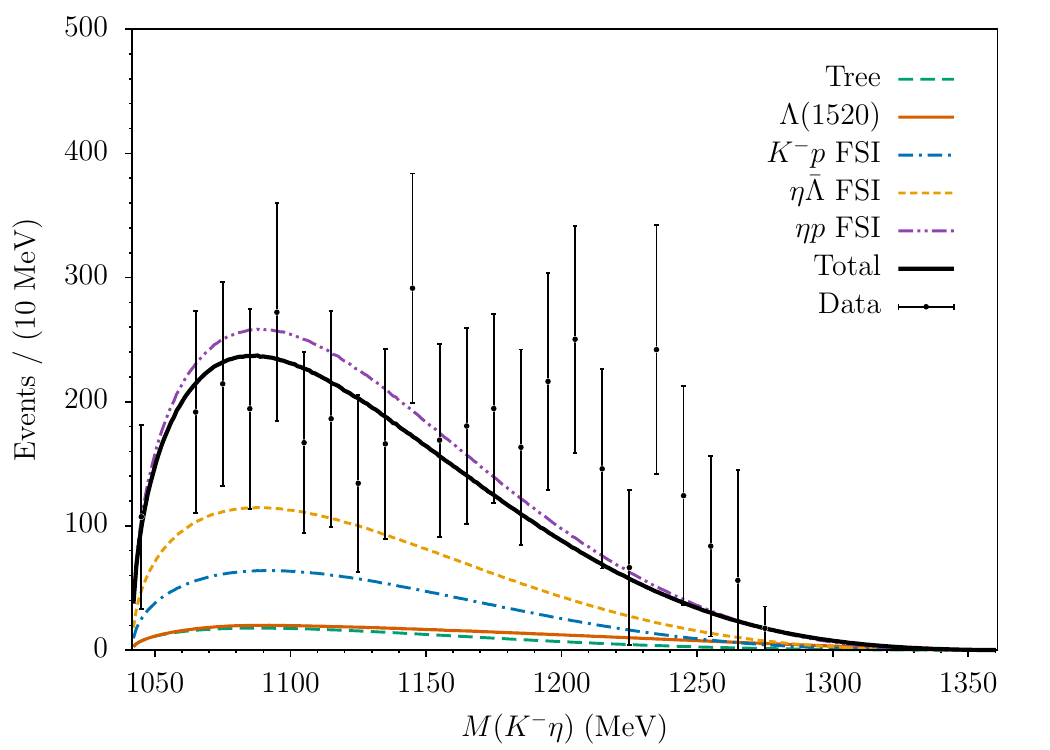}
		\label{fig:fit_Kmeta}
	}\hfill
	\subfigure[]{
		\includegraphics[width=0.32\textwidth]{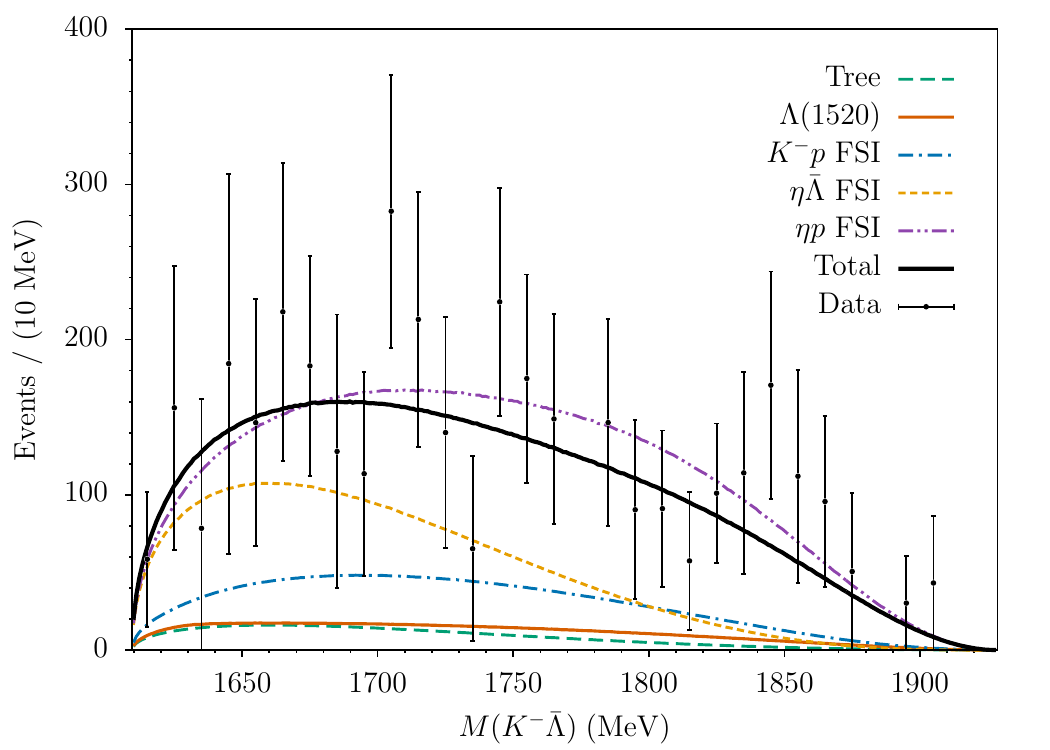}
		\label{fig:fit_KmLambar}
	}\hfill
	\subfigure[]{
		\includegraphics[width=0.32\textwidth]{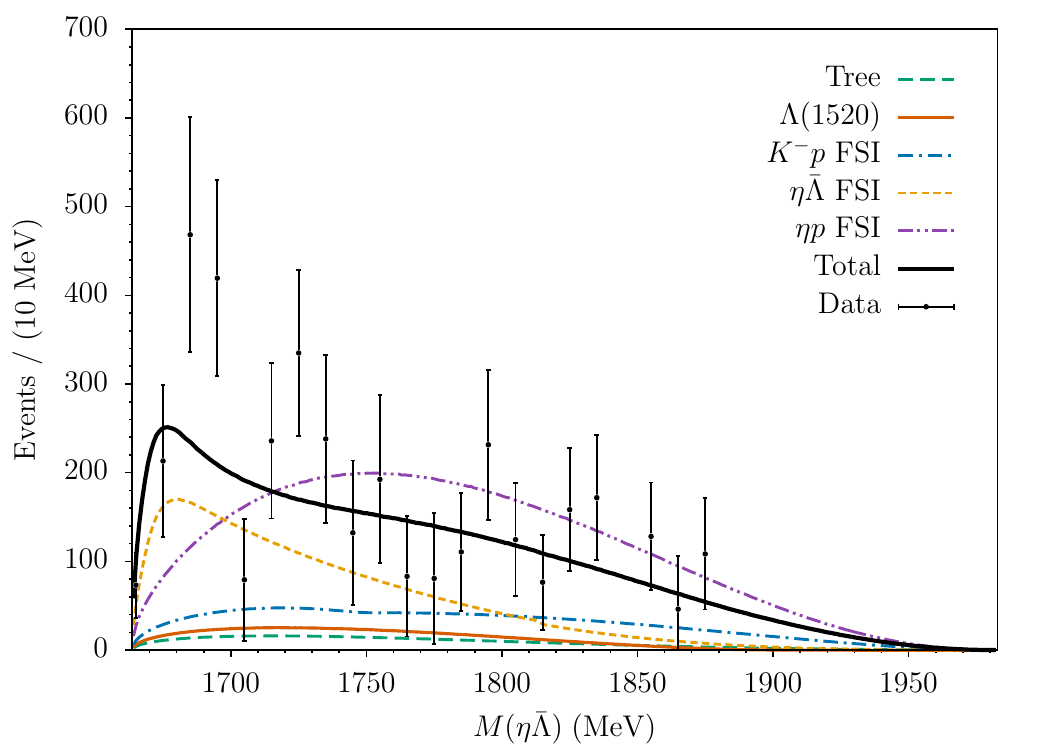}
		\label{fig:fit_etaLambar}
	}
	
	\caption{Invariant mass distributions of $pK^-$~(a), $p\eta$ (b), $p\bar{\Lambda}$ (c), $K^-\eta$ (d), $K^-\bar{\Lambda}$ (e), and $\eta\bar{\Lambda}$ (f) compared with the efficiency-corrected experimental data provided by the BESIII Collaboration~\cite{BESIII:2026gro,weiminsong}.}
	\label{fig:invariant_mass_distribution}
\end{figure*}

\begin{figure*}[htbp]
	\centering
	
	\subfigure[]{
		\includegraphics[width=0.32\textwidth]{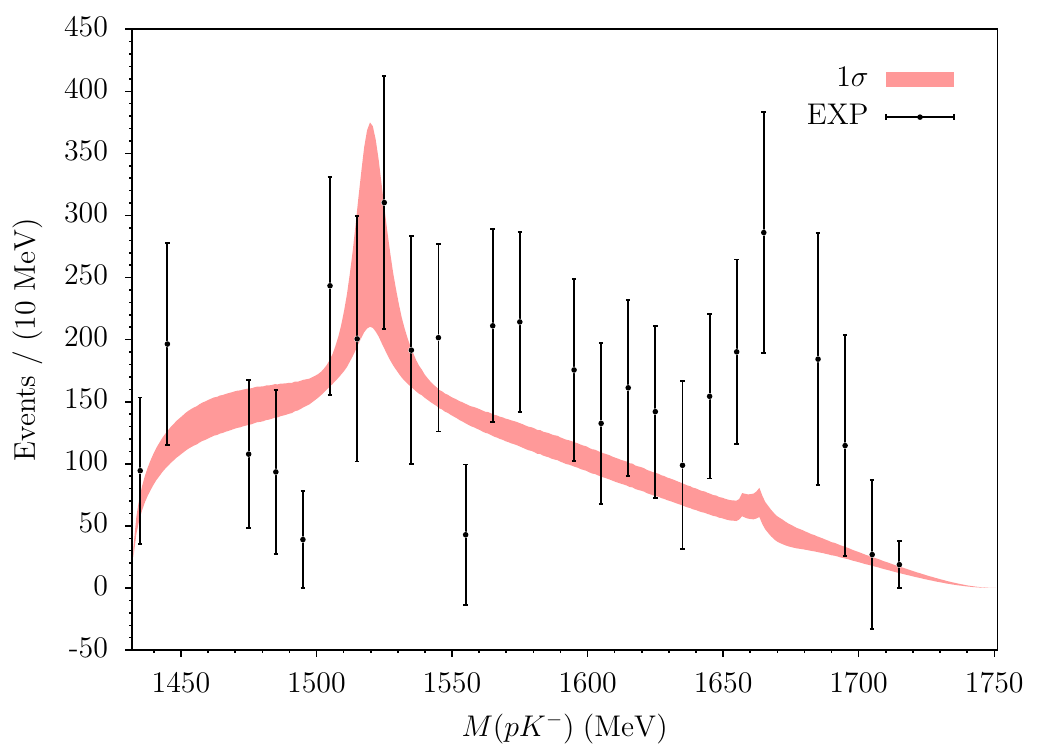}
		\label{fig:error_pKm}
	}\hfill
	\subfigure[]{
		\includegraphics[width=0.32\textwidth]{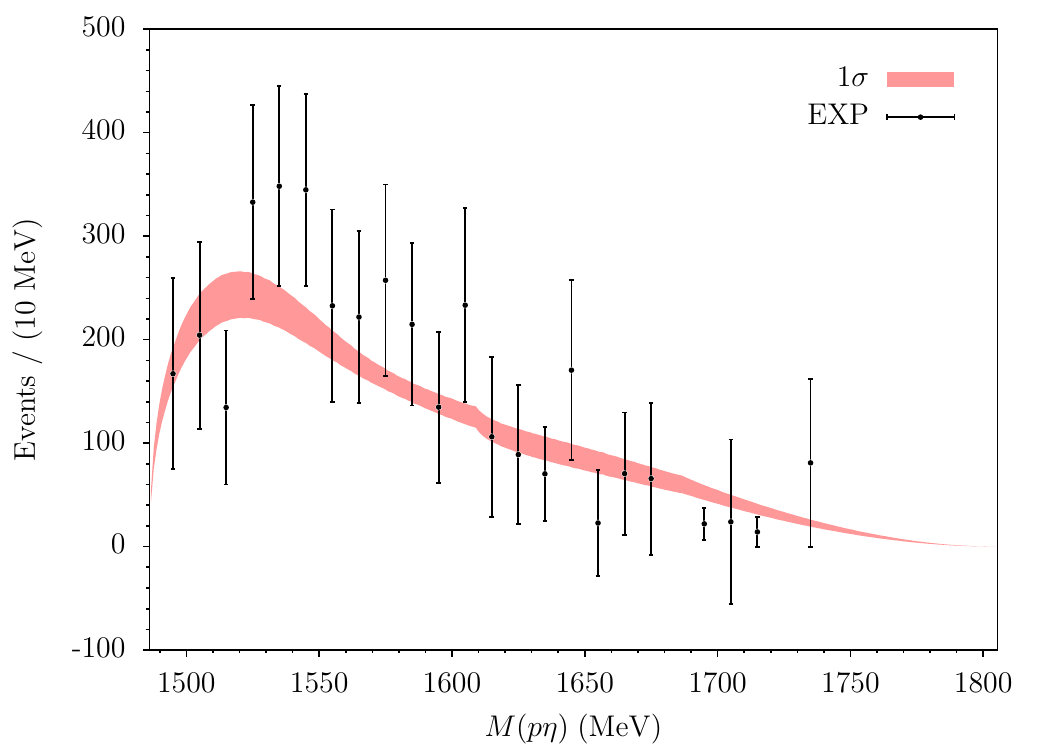}
		\label{fig:error_peta}
	}\hfill
	\subfigure[]{
		\includegraphics[width=0.32\textwidth]{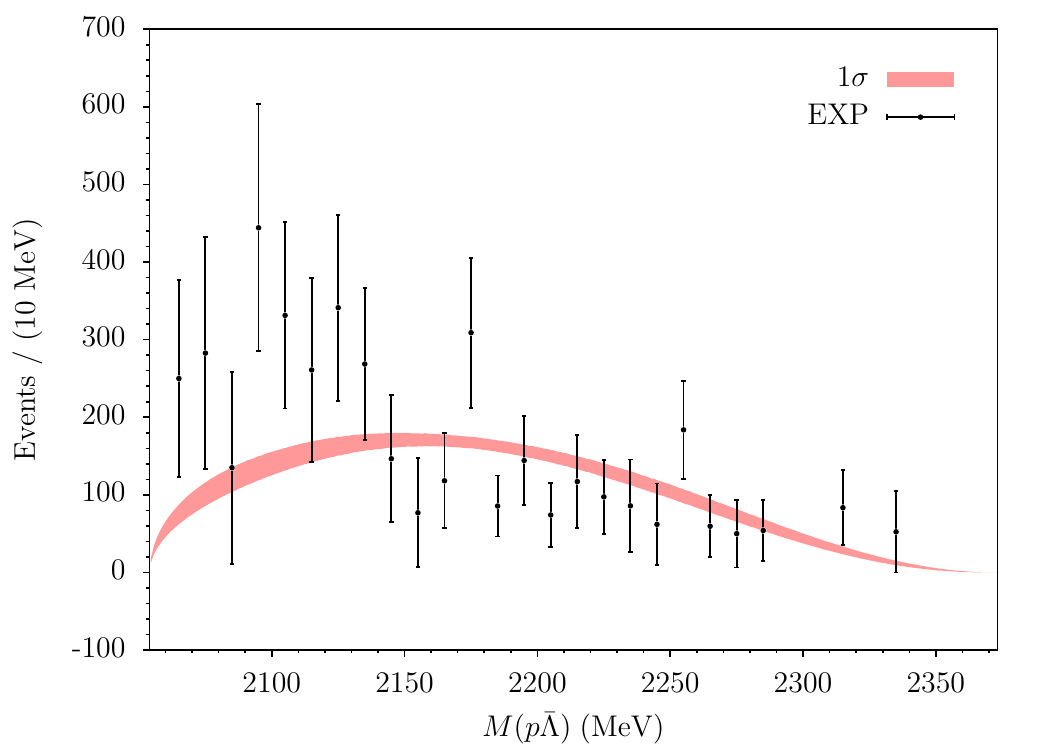}
		\label{fig:error_pLambar}
	}
	
	\par\vspace{0.3cm}

	\subfigure[]{
		\includegraphics[width=0.32\textwidth]{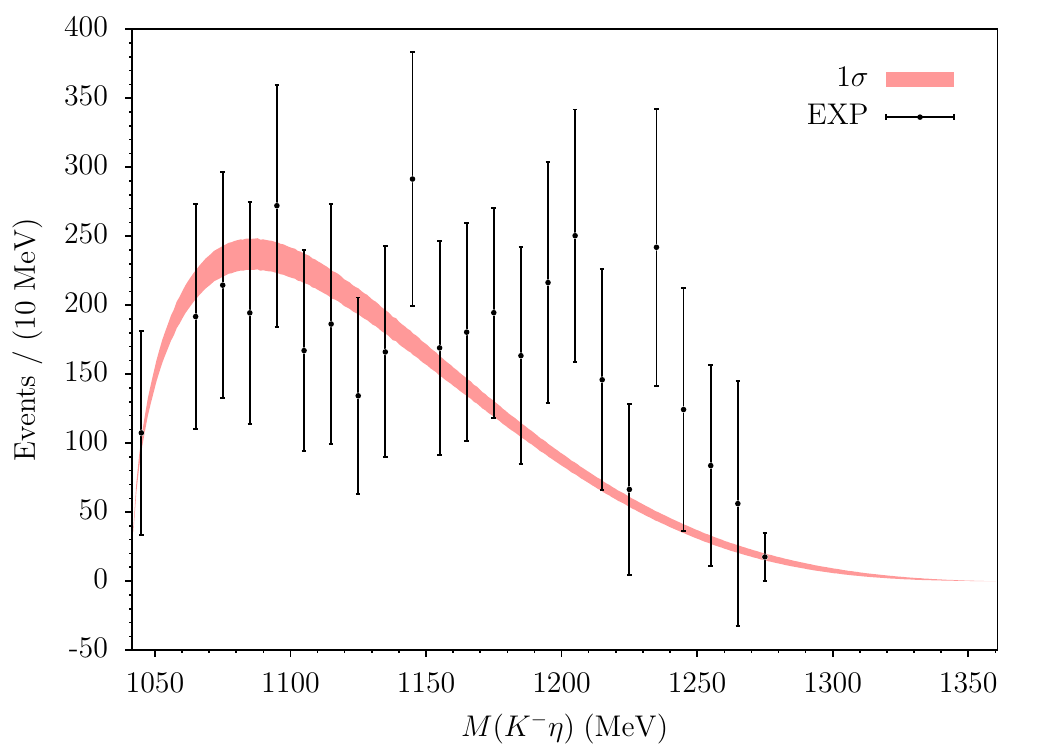}
		\label{fig:error_Kmeta}
	}\hfill
	\subfigure[]{
		\includegraphics[width=0.32\textwidth]{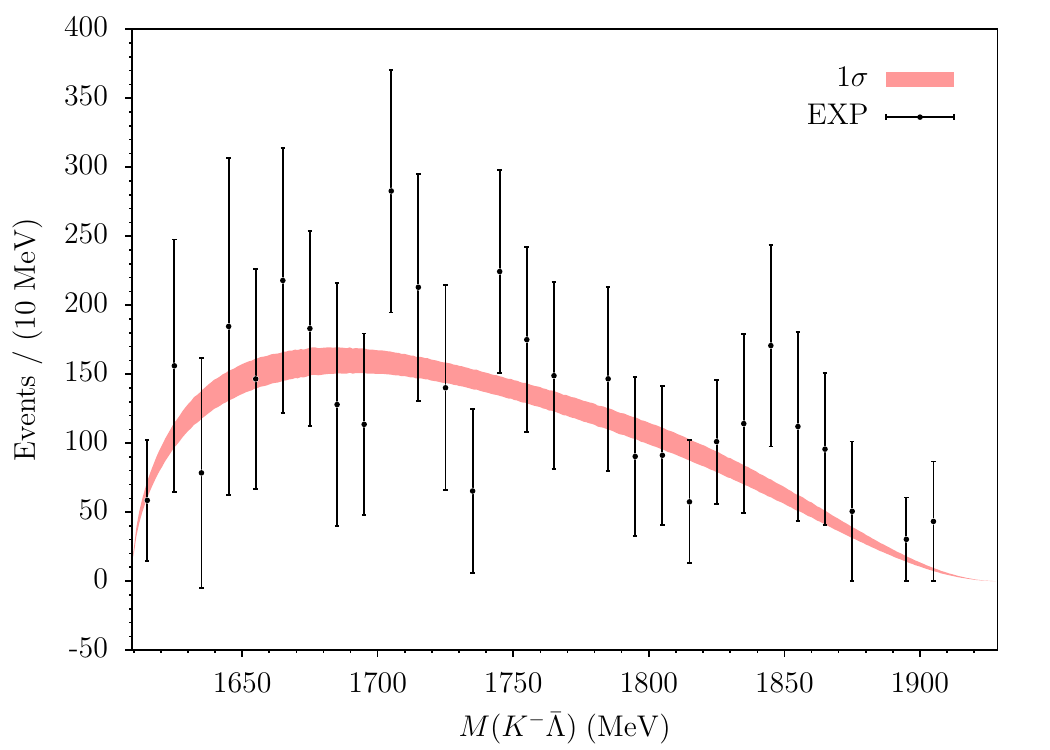}
		\label{fig:error_KmLambar}
	}\hfill
	\subfigure[]{
		\includegraphics[width=0.32\textwidth]{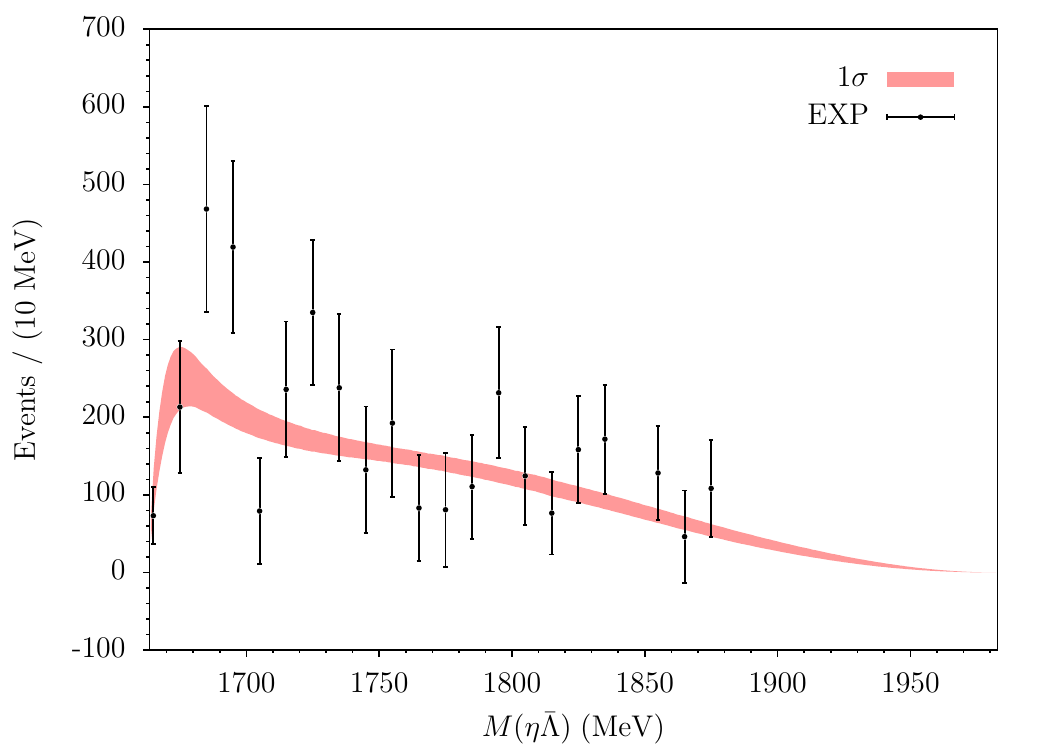}
		\label{fig:error_etaLambar}
	}
	
	\caption{Theoretical uncertainties in the invariant mass distributions of $pK^-$ (a), $p\eta$ (b), $p\bar{\Lambda}$ (c), $K^-\eta$ (d), $K^-\bar{\Lambda}$ (e), and $\eta\bar{\Lambda}$ (f), estimated using the parametric bootstrap method. The efficiency-corrected experimental data are provided by the BESIII Collaboration~\cite{BESIII:2026gro,weiminsong}.}
	\label{fig:error_mass_distribution}
\end{figure*}

Since the coefficients $A$, $B$, and $C$ in Eq.~\eqref{Eq:Trace} need not be real, we allow them to be complex by introducing two relative phases, $\phi_A$ and $\phi_C$, such that $A=|A|e^{i\phi_A}$ and $C=|C|e^{i\phi_C}$. Our model therefore contains six free parameters, namely $|A|$, $B$, $|C|$, $D$, $\phi_A$, and $\phi_C$. A fit to the efficiency-corrected invariant-mass distributions provided by the BESIII Collaboration~\cite{BESIII:2026gro,weiminsong} yields $|A|=(17.85\pm2.43)\times10^{-2}$, $B=(1.54\pm0.36)\times10^{-2}$, $|C|=(10.67\pm2.27)\times10^{-2}$, $D=(0.63\pm0.21)\times10^{-2}$, $\phi_A=(-28.43\pm7.33)^\circ$, and $\phi_C=(131.43\pm9.42)^\circ$, with $\chi^2/\mathrm{d.o.f.}=141.68/141=1.00$. The resulting fit provides a good overall description of the efficiency-corrected distributions.

Figure~\ref{fig:invariant_mass_distribution} shows the invariant mass distributions of $pK^-$, $p\eta$, $p\bar{\Lambda}$, $K^-\eta$, $K^-\bar{\Lambda}$, and $\eta\bar{\Lambda}$. The green dashed curves denote the tree-level contributions, whereas the red solid curves show the contributions from the intermediate $\Lambda(1520)$ state. The blue dash-dotted, yellow dashed, and purple dash-dotted curves correspond to the $K^-p$, $\eta\bar{\Lambda}$, and $\eta p$ FSI contributions, respectively, given in Eqs.~\eqref{Eq:t_k-p}, \eqref{Eq:t_etaLambdabar}, and \eqref{Eq:t_etap}. Finally, the black solid curves show the full results obtained from the total amplitude in Eq.~\eqref{Eq:ttotal}. While the individual curves show the distributions obtained from each mechanism separately, whereas the full distributions also include interference among the tree-level and FSI amplitudes. However, interference between the amplitude associated with the intermediate $\Lambda(1520)$ resonance and the remaining amplitudes is neglected, as specified in the formalism.

The $K^-p$ invariant mass distribution in Fig.~\ref{fig:fit_pKm} is dominated by the pronounced peak near $1520$~MeV, which originates from the intermediate $\Lambda(1520)$ state. Its position and narrow line shape are consistent with the prominent structure observed in the experimental distribution. In addition to this peak, the $K^-p$ FSI contribution exhibits a near-threshold enhancement associated with the higher-mass pole of the $\Lambda(1405)$ in our coupled-channel model. Furthermore, it displays a weaker cusp-like feature around 1670~MeV associated with the $\Lambda(1670)$. These FSI effects are more readily identified in the separate $K^-p$ FSI contribution than in the full distribution.

In the $p\eta$ invariant mass distribution shown in Fig.~\ref{fig:fit_peta}, the isolated $\eta p$ FSI contribution displays a broad enhancement around $1535$~MeV. Within our model, this behavior is associated with the dynamically generated $N(1535)$ state. Meanwhile, the $\eta\bar{\Lambda}$ distribution in Fig.~\ref{fig:fit_etaLambar} shows a pronounced near-threshold enhancement in the corresponding FSI contribution. Within the coupled-channel model, this structure is associated with the dynamically generated $\bar{\Lambda}(1670)$ state.

In this context, it is instructive to compare the manifestations of the $\Lambda(1670)$ and its antiparticle in the $K^-p$ and $\eta\bar{\Lambda}$ invariant mass distributions. The two FSI amplitudes are calculated using the same strong-interaction parameters, yet their resonance signals differ considerably. For the adopted parameters, the pole position is $1669.11-19.62\,i$~MeV, and the squared coupling magnitudes are $|g_{K^-p}|^2=0.05$ and $|g_{\eta\Lambda}|^2=1.91$. The larger coupling to $\eta\Lambda$ helps explain why the near-threshold signal is more pronounced in the $\eta\bar{\Lambda}$ channel than in the $K^-p$ channel. 

Although we have obtained the efficiency-corrected data provided by the BESIII Collaboration, the uncertainties in the current experimental data still limit how firmly these local structures can be established. For example, in the $K^-p$ distribution, the bin centered at $1665$~MeV contains a relatively large number of events, whereas the neighboring bin at $1675$~MeV has a central value of zero. Such bin-to-bin fluctuations caution against drawing strong conclusions from individual points. Higher-statistics measurements would therefore be valuable for testing the predicted threshold behavior and constraining the relative FSI strengths.  In particular, future measurements at Belle~II and the proposed Super Tau-Charm Facility (STCF)~\cite{Achasov:2023gey} could provide important complementary information.

To estimate the theoretical uncertainties and assess the sensitivity
of the model predictions to the experimental data, we employ a parametric bootstrap~(resampling) procedure~\cite{Efron:1986hys,Albaladejo:2016hae,Lyu:2026rsm}. Specifically, we generate $10^6$ pseudo-data sets by drawing each pseudo-data point from a Gaussian distribution centered on the corresponding measured value, with a standard deviation equal to the quoted experimental uncertainty. The original uncertainties are retained when fitting each pseudo-data set. We then refit the six model parameters and calculate the corresponding invariant mass distributions for each fitted parameter set. The resulting uncertainty bands are shown in Fig.~\ref{fig:error_mass_distribution} and represent the $1\sigma$ statistical uncertainty propagated through this resampling procedure. Thus, they quantify the sensitivity of the fitted distributions to fluctuations in the experimental input within the assumed model.

\section{ Conclusions }\label{sec4}

In this work, motivated by the recent BESIII measurement, we have investigated the $\chi_{c0}\to pK^-\bar{\Lambda}\eta$ decay within the chiral unitary approach. The primary production vertex is constructed from three SU(3)-invariant single-trace structures, and the subsequent $S$-wave final-state interactions in the $K^-p$, $\eta\bar{\Lambda}$, and $\eta p$ subsystems are incorporated through coupled-channel amplitudes. These amplitudes dynamically generate the $\Lambda(1405)$, $\Lambda(1670)$, and $N(1535)$ resonances, while the intermediate $\Lambda(1520)$ contribution is included explicitly. With six free parameters, our model provides a reasonable overall description of all six measured two-body invariant mass distributions, yielding $\chi^2/\mathrm{d.o.f.}=1.00$.

The pronounced peak in the $K^-p$ distribution is accounted for by the $\Lambda(1520)$ contribution. The separate $K^-p$ FSI contribution also exhibits a near-threshold enhancement associated with the higher-mass pole of the two-pole structure of the $\Lambda(1405)$ and a weaker cusp-like feature associated with the $\Lambda(1670)$. In contrast, the $\eta\bar{\Lambda}$ FSI contribution displays a clear near-threshold enhancement associated with the $\bar{\Lambda}(1670)$. This difference reflects the stronger coupling of the $\Lambda(1670)$ to $\eta\Lambda$ than to $K^-p$, together with the effects of the production mechanisms and their interference. The isolated $\eta p$ FSI contribution shows a broad enhancement associated with the $N(1535)$.

We have estimated the theoretical uncertainties of the predicted invariant mass distributions using a resampling approach, which quantifies their sensitivity to fluctuations in the experimental input within the assumed model. The current experimental uncertainties limit firm conclusions about the less prominent resonance-related and threshold structures. Higher-precision measurements at BESIII, Belle~II, and the proposed STCF would provide valuable tests of the predicted line shapes and tighter constraints on the relative FSI strengths. The $\chi_{c0}\to pK^-\bar{\Lambda}\eta$ decay therefore offers a useful setting for investigating coupled-channel dynamics and further clarifying the properties of low-lying baryon resonances.

\section*{Acknowledgments}
We acknowledge Prof. Eulogio Oset for fruitful discussions, and Prof. Wei-Min Song for providing the efficiency-corrected experimental data used in this work. This work was supported by the National Key R\&D Program of China (Grant No. 2024YFE0105200), the National Natural Science Foundation of China under Grant No. 12475086 and No. 12192263, the Natural Science Foundation of Henan (Grant No. 252300423951), and the Zhengzhou University Young Student Basic Research Projects for PhD students (Grant No. ZDBJ202522). Wen-Tao Lyu acknowledges the support of the China Scholarship Council.

\end{document}